\documentstyle[amstex,preprint,aps,epsfig]{revtex}
\tightenlines
\newcommand{\beq}{\begin{equation}}
\newcommand{\eeq}{\end{equation}}
\newcommand{\beqa}{\begin{eqnarray}}
\newcommand{\eeqa}{\end{eqnarray}}

\begin{document}
\title{Pseudogaps due to sound modes: from incommensurate charge density waves to
semiconducting wires.}
\author{S. A. Brazovskii$^{1,2}$ and S.I. Matveenko$^{2,1}$}
\address{$^{1}$Laboratoire de Physique Th\'{e}orique et des Mod\`{e}le Statistiques,\\
CNRS, B\^{a}t.100, Universit\'{e} Paris-Sud, 91405 Orsay cedex, France. \\
$^{2}$L.D. Landau Institute for Theoretical Physics,\\
Kosygina Str. 2, 117940, Moscow, Russia.}
\date{06/08/02}
\draft
\maketitle
\begin{abstract}
We consider pseudogap effects for electrons interacting with gapless modes.
We study both generic 1D semiconductors with acoustic phonons and
incommensurate charge density waves. We calculate the subgap absorption as
it can be observed by means of the photo electron or tunneling spectroscopy.
Within the formalism of functional integration and the adiabatic
approximation, the probabilities are described by nonlinear configurations
of an instanton type. Particularities of both cases are determined by the
topological nature of stationary excited states (acoustic polarons or
amplitude solitons) and by presence of gapless phonons which change the
usual dynamics to the regime of the quantum dissipation. Below the free
particle edge the pseudogap starts with the exponential (stretched
exponential for gapful phonons) decrease of transition rates. Deeply within
the pseudogap they are dominated by a power law, in contrast with nearly
exponential law for gapful modes.
\end{abstract}


\section{Introduction: pseudogaps in 1D.}

This article is devoted to theory of pseudogaps (PGs) in{\em \ }electronic
spectras in applications to Photo Electron Spectroscopy (PES). We shall
study an influence of quantum lattice fluctuations upon electronic
transitions in the subgap region for one-dimensional ($1D$) systems with
gapless phonons. Low symmetry systems with gapful spectra have been
addressed by the authors recently \cite{matv-01} and we refer to this
article for a more comprehensive review and references. Here we will show
that sound branches of phonon spectra change drastically the transition
rates making them much more pronounced deeply within the PG. We shall
consider two types of systems: generic $1D$ semiconductors with acoustic
e-ph coupling (conducting polymers, quantum wires, nanotubes) and
Incommensurate Charge Density Waves (ICDWs) \cite{gruner} which possess the
gapless collective phase mode.

The PG concept \cite{lra} refers to various systems where a gap in their
bare electronic spectra is partly filled showing subgap tails. Even for pure
systems and at temperature $T=0$ there may be a rather smeared edge
$E_{g}^{0}$ while the spectrum extends deeply inwards the gap till some
absolute edge $E_{g}$ which may be even zero (no true gap at all). A most
general reason is that stationary excitations (eigenstates of the total
electron-phonon $e-ph$ system) are the self-trapped states, {\em polarons}
or {\em solitons}, which energies, $W_{p}$ or $W_{s}$, are below the ones of
free electrons thus forming the absolute edge at $E_{g}<E_{g}^{0}$.
Nonstationary states filling the PG range $E_{g}^{0}>E>E_{g}$ can be
observed {\em only via instantaneous measurements} like optics, PES or
tunneling. Particularly near $E_{g}^{0}$ the states resemble free electrons
in the field of {\em uncorrelated quantum fluctuations} of the lattice \cite
{braz-76}; here the self-trapping has not enough time to be developed. But
approaching the exact threshold $E_{g}$, the excitations evolve towards
eigenstates which are {\em self-trapped} $e-ph$ complexes. The PGs must be
common in $1D$ semiconductors just because of favorable conditions for the
self-trapping \cite{rashba}. Further on, the PG is especially pronounced
when the bare gap is opened spontaneously as a symmetry breaking effect. In
quasi-$1D$ conductors it is known as the Peierls-Fr\"{o}hlich instability
leading to the CDW formation. Here the picture of the PG has been suggested
first theoretically \cite{lra} (recall also \cite{barisic} and another model 
\cite{sadovskii}) in relation to absence of the long range order in $1D$
CDWs at finite temperature. In this approach the smearing of the
mean field electronic gap $2\Delta _{0}$ corresponds to
disappearance of the true Peierls-Fr\"{o}hlich transition in favor of a
smooth crossover. The PG shape was related, and derived from, the
temperature dependent finite correlation length $\xi $. An alternative
picture was suggested in \cite{braz-76} and further developed in \cite
{braz-80}. It concentrates on effects persistent even at zero temperature
which are due to strong interaction between bare electronic excitations and
perturbations (amplitude and phase phonons) of the CDW ground state. Here
the PG in instantaneous electronic spectra becomes related to transformation
of electrons into solitons. 

Experimentally, the PGs in ICDWs were addressed
first by optic \cite{optics,nad-opt,brill} and more recently by the PES and
ARPES \cite{pes}. The theoretical interpretation of earlier
experiments was done in \cite{ohio} by compilation of approaches from \cite
{lra,braz-76,sadovskii}. Detailed theories of the subgap absorption in
optics have been developed already for systems with low symmetries
(nondegenerate, like semiconductors with gapful phonons, or discretely
degenerate like the dimerized Peierls state). They addressed first a general
type of polaronic semiconductors \cite{iosel} with emphasis to long range
Coulomb effects, and the $1D$ Peierls system emphasizing solitonic
processes \cite{kiv-86}. Recently the authors \cite{matv-01} extended the
theory of pseudogaps to single electronic spectra in application to the PES
and, particularly intriguing, to the ARPES (momentum resolved PES) probes.
But properties of ICDWs are further complicated by appearance of the gapless
collective mode which bring drastic changes. The case of acoustic polarons
(APs) in a 1D semiconductor belongs to the same class while it is not
usually noticed.

Specifics of $1D$ systems with continuous degeneracy (with respect to the
phase for the ICDW, to displacements for usual crystals) is that even single
electronic processes can create topologically nontrivial excitations, the
solitons. Thus for the ICDW a single electron or hole with energy near the
gap edges $\pm \Delta _{0}$ spontaneously evolve to the nearly amplitude
soliton - AS whilst the original particle is trapped at the local level near
the gap center. The energy $\approx 0.3\Delta _{0}$ is released, at first
sight within a time $\omega _{ph}^{-1}$ $\sim 10^{-12}s$. We will see that
actually there is also a long scale adaptation process which determines
shapes of transition probabilities. Similarly, the usual acoustic polaron in 
a 1D semiconductors is characterized by the electronic density $\rho $
selflocalized within the potential well $\sim \partial \varphi /\partial
x\sim \rho $, hence a finite increment
$\varphi(+\infty) -\varphi(-\infty)\sim\int\rho dx $ 
of the lattice displacements $\varphi $ over the length $x$ which 
is the signature of topologically nontrivial
solitons. These systems with continuous degeneracy form a special class
which shows particular properties and must be studied differently than in 
\cite{matv-01}. They are addressed in this article.

\section{Functional integrals and instantons for the PES.\label{general}}

The absorption rate 
$ I(\Omega,P )$, as a function of frequency $(\Omega $ and momentum $P$, 
for the ARPES can be expressed in terms of the
spectral density of the one-electron retarded Green function
$G(t,t^{\prime};x,x^{\prime })$ as
\begin{equation}
I(P,\Omega )\propto Im\int dXe^{-iPX}\int_{0}^{\infty }dTe^{i\Omega
T}G(X,T,0,0).  \label{arpes}
\end{equation}
We shall address here the simple PES, non resolved in momenta, measures the integrated absorption
intensity $I(\Omega )=\int I(P,\Omega )dP/2\pi $. (Since now on we shall
omit all constant factors and take the Plank constant $\hbar =1$; $\Omega $
will be measured with respect to a convenient level: the band edge for
semiconductors or the middle of the gap for CDWs.)

We shall use the adiabatic approximation valid when changes of electronic
energies are much larger than relevant phonon frequencies. Electrons are
moving in the slowly varying phonon potential, e.g. ${\rm Re}\{\Delta
(x,t)\exp [i2k_{F}x]\}$ for the ICDW, so that at any instance $t$ their
energies $E(t)$ and wave functions $\psi (x,t)$ are defined as eigenstates
for the instantaneous lattice configuration and they depend on time only
parametrically. In the following we shall work in the Euclidean space
$it\rightarrow t$ which is adequate for studies of classically forbidden
processes \cite{iosel,raja,iord-rash}. Then in the form of functional
integrals over lattice configurations we have 
\begin{equation}
I(\Omega )\propto \int_{0}^{\infty }dT\int D[\Delta (x,t)]\psi _{0}(0,T)\psi
_{0}^{+}(0,0)e^{-S},  \label{I}
\end{equation}
Here $\psi _{0}$ is the wave function of the particle (for the PES it is
actually a hole) added and extracted at moments $0$ and $T$. For
calculations of subgap processes only the lowest singly filled localized
state is relevant which energy $E_{0}$ is split off inside the gap. The
action $S=S[\Delta (x,t),T]$ 
\begin{equation}
S={(\int_{-\infty }^{0}+\int_{T}^{\infty
})dt\,L_{0}+\int_{0}^{T}dt\,(L_{1}-\Omega )},\;L_{1}-L_{0}=E_{0}  \label{S-L}
\end{equation}
is given by Lagrangians $L_{j}[\Delta ]$ where the labels $j=0,1$ correspond
to ground states for $2M$ (the bare number) and $2M\pm 1$ electrons in the
potential $\Delta (x,t)$. The main contribution comes from saddle points of
$S$, the instantons, which are extremas with respect to both the function
$\Delta (x,t)$ and the time $T$. There are also special cases \cite{matv-01}, 
particularly important for the ARPES, when the
extremum must be taken for the whole expression in the integral (\ref{I}),
taking into account the wave functions in the prefactor. Otherwise we obtain
for the stationary point $dS/dT=0$, that is $E_{0}(0)=E_{0}(T)=\Omega $
which determines $T(\Omega )$.

In the following we shall concentrate on most principle features leaving
aside calculations of prefactors and the question of the momentum dependence
necessary for the ARPES. For a simpler case of nondegenerate systems they
have been studied in \cite{matv-01}.

\section{Creation of amplitude solitons in ICDWs.} \label{soliton}

Consider first the subgap electronic spectra for the ICDW described by the
Peierls-Fr\"{o}hlich model. The ICDW order parameter is the complex field
$\Delta =|\Delta (x,t)|\exp [i\varphi (x,t)]\,$ acting upon electrons by
mixing states near the Fermi momenta points $\pm k_{F}$. The Lagrangians
$L_{j}$ consist of the bare kinetic $\sim |\partial _{t}\Delta |^{2}$ and
potential $\sim |\Delta |^{2}$ lattice energies and of the sum over the
filled electron levels, in the $j$-th state:
\[
L_{j}=\int dx2|\partial _{t}\Delta |^{2}/(\pi v_{F}\omega
_{0}^{2})+V_{j}[\Delta (x,t)]\, 
\]
where $v_{F}$ is the Fermi velocity in the metallic state and $\omega {0}$
is the amplitude mode frequency. ($\omega _{0}\ll \Delta _{0}$ is the
condition for the adiabatic approximation.) 

The important fact is that the stationary state of the system with an
odd number of particles, the minimum of $V_{1}$, is the amplitude soliton
($AS$) $\Delta \Rightarrow -\Delta $ with the midgap state $E_{0}=0$ occupied
by the singe electron. Evolution of the free electron with the initial
$E_{0}=\Delta _{0}$ to the AS with $W_{s}=2/\pi \Delta _{0}<\Delta _{0}$ can
be fortunately described by the known exact solution for intermediate
configurations characterized by the single
intragap $E_{0}=\Delta _{0}\cos \theta $ with $0\leq \theta \leq \pi$,
 hence $-\Delta _{0}\leq E_{0}\leq \Delta _{0}$. It was found \cite{braz-80}, 
see also reviews \cite{braz-84,braz-89}, to be the Chordus Soliton ($ChS$)
with $2\theta $ as the total chiral angle: $\Delta (+\infty )/\Delta
(-\infty )=\exp (2i\theta )$, see Fig.1 and the Appendix for
details. The filling number of the intragap state $\nu =0,1$ corresponds to
labels $j=0,1$. The term $V_{0}(\theta )$ increases monotonically from
$V_{0}(0)=0$ for the $2M$ GS to $V_{0}(\pi )=2\Delta _{0}$ for the
$2M+2$ GS with two free holes. The term $V_{1}(\theta )=V_{1}(\pi -\theta )$
is symmetric describing both the particle upon the $2M$ GS and the hole upon
the $2M+2$ GS. Apparently $V_{1}(0)=V_{1}(\pi )=\Delta _{0}$ while the
minimum is reached at $\theta =\pi /2$ that is for the purely AS: $\min
V(\theta )=$ $V_{1}(\pi /2)=W_{s}<\Delta _{0}$ where $W_{s}=2\Delta _{0}/\pi$ 
is the AS energy, see Fig.2. Thus, to create a nearly AS with $\theta
=90^{\circ }$,
 the light with $\Omega \approx W_{s}$ is absorbed by the
quantum fluctuation
 with $E_{0}(\theta )=W_{s}$ which is close to the chordus
soliton with the
 angle $\theta \approx 50^{\circ}$.

Notice that being the uncharged spin carrier with the topological charge
equal unity, the AS is a quasiclassical realization of the {\it spinon} in
systems with nonretarded attraction of electrons (that is with high, rather
than low, phonon frequencies). Thus our analysis is applied qualitatively
also to arbitrary nonadiabatic electronic systems provided that they are
found in the spin gap regime. (See also the next chapter.)

Usually it is tempting to use the static solution, with some free parameter,
as an Anzatz for the time dependent process which proved to be successful in
gapful cases \cite{matv-01,kiv-86}. But here, putting $\theta \rightarrow
\theta (t)$, we would arrive at $\partial _{t}\Delta \neq 0$ at all $x$
hence the action would be infinite $S\sim $ {\it the system length}. The
vanishing probability simply reflects the fact that a globally finite
perturbation, characteristic for topologically nontrivial solitons, cannot
spread over the whole length by a finite time. More generally, as a
topologically nontrivial object, the AS cannot be created in a pure form:
adaptational deformations must appear to compensate for the topological
charge. These deformations are developing over long space-time scales which
can be described in terms of the gapless mode, the phase $\varphi $, alone.
Hence allowing for the time evolution of the chiral angle $\theta
\rightarrow \theta (t)$ within the core, we should also unhinder the field
$\varphi \rightarrow \varphi (x,t)$ at al $x$ and $t$. The resulting
trajectory is shown at the Fig.1 for an instant of time. Starting
from $x\rightarrow -\infty $ and returning to $x\rightarrow \infty $ the
configuration follows closely the circle $|\Delta |=\Delta _{0}$ changing almost
entirely by phase. Approaching the soliton core the phase matches
approximately the angles $\pm \theta $ which delimit the chordus part of the
trajectory. The whole trajectory is closed which allows for the finite
action.

Except for a short time scale $T<\xi _{0}/u$ (see Ch.\ref{noise})
characterized by small $\theta $ and large lengths
$\xi =\xi _{0}/\sin\theta $, the configuration $\Delta (x,t)$ can be divided
into the inner part, the core at $|x|\sim \xi $, and the outer part
$|x|\gg \xi $ where only perturbations of the phase $\varphi (x,t)$
are important. The inner
part can be described by the chordus soliton $\Delta _{ChS}(x,t)$. The
chordus angle $2\theta (t)$ evolves in time from $\theta (\pm \infty )=0$ to 
$\theta _{m}$ in the middle of the $T$ interval. For $T\rightarrow \infty $,
that is near the stationary state of the AS,
$\theta _{m}\rightarrow \pi /2$. Actually this value is preserved during most 
of the $T$ interval so that
changes between $\theta =0$ and $\theta =\pi /2$ are concentrated within
finite ranges $\tau _{0}\sim \xi _{0}/u\ll T$ near the termination points.
From large scales we view only a jump
$\varphi (x,t)\approx \theta (t){\rm sgn}(x)$ with
$\theta (t)\approx \theta _{m}\Theta (t)\Theta (T-t)$ ($\Theta$ is the standard step function). 
Since the configuration stays close to
the AS during the time $T$, the main core contribution to the action is 
\begin{equation}
S_{core}=(W_{s}-\Omega )T+\delta S_{core}  \label{S-core}
\end{equation}
where the first correction $\delta S_{core}^{0}=cnst$ comes form regions
around moments $0,T$ independently. The significant $T$ dependent contribution 
$\delta S(T)$ comes from interference of regions $0$ and $T$. Their
interaction via gapful excitations like the amplitude mode decays
exponentially as $\delta S_{gap}\sim \exp (-\omega _{0}T)$ and for low
symmetry systems there was no other contribution. But now, for the ICDW,
there are {\em sound modes providing the main effect} which is addressed
below.

Matching the inner and outer regions is not well defined unless we consider
a full microscopic time dependent model which is not possible. But
fortunately the long range effects can be treated easily if we generalize
the scheme suggested earlier for static problems of solitons at presence of interchain interactions 
\cite{soliton3d,braz-80}. The outer region is described by the action for
the sound like phase mode 
\begin{equation}
S_{snd}[\varphi (x,t),\theta (t)]=\frac{v_{F}}{4\pi }
\int \int dxdt\left((\partial _{t}\varphi /u)^{2}+
(\partial _{x}\varphi )^{2}\right)
\,,\;\,\varphi (t,x_{s}\pm 0)=\mp \theta (t)  \label{s-snd}
\end{equation}
where $u$ is the phase velocity. The action $S_{snd}$ takes into account the
source provided by the chordus soliton forming around $x_{s}$ which enforces
the discontinuity of $2\theta $. Integrating out $\varphi (x,t)$ from
$\exp\{-S_{snd}[\varphi ,\theta ]\}$ with this condition we arrive at the action
for $\theta (t)$ 
\begin{equation}
S_{snd}[\theta ]\approx \frac{v_{F}/u}{2\pi ^{2}}
\int \int dt_{1,2}\dot{\theta}(t_{1})\ln |(t_{1}-t_{2})|\dot{\theta}(t_{2})=
\frac{v_{F}/u}{2\pi ^{2}}
\int \int dt_{1,2}\left( \frac{\theta (t_{1})-
\theta (t_{2})}{t_{1}-t_{2}}\right) ^{2}  \label{dis}
\end{equation}
Here the last form is a typical action for the problem of quantum
dissipation \cite{leggett} $S\sim \sum |\omega ||\theta _{\omega }|^{2}$. In
our case this dissipation comes from emission of phase phonons while forming
the long range tail in the course of the chordus soliton development.
This action, together with $V_j$, can be used to prove the above statements on the time evolution of the ChS core. 

Remember now that $\dot{\theta}=\partial _{t}\theta $ is peaked within
narrow regions $\sim \xi _{0}/u$ around moments $t=0$ and $T=0$ and close to
zero otherwise. Then 
\begin{equation}
S_{snd}\approx \left( v_{F}/4u\right) \ln (uT/\xi _{0})  \label{log}
\end{equation}

There is an even more phenomenological point of view, see \cite{comb-top} on
more details and examples for combined topological defects. The AS creates
the $\pi -$ discontinuity along its world line $(0<t<T,0)$. To be
topologically allowed, that is to have a finite action, the line must
terminate with half integer vortices located at $(0,0)$ and $(0,T)$ which
circulation will provide the compensating jump $\delta \varphi =\pi $ along
the interval ($\Delta \Rightarrow -\Delta $ combined with
$\varphi\Rightarrow \varphi +\pi $ leaves the order parameter
$\Delta \exp (i\varphi)$ invariant). Then the standard energy of vortices for (\ref{s-snd}) leads
to the action (\ref{log}). Contrary to usual $2\pi -$ vortices, the line
connecting the half-integer ones is the physical singularity which tension
gives (\ref{S-core}).

Minimizing $S_{tot}=S_{core}+S_{snd}\,$with respect to $T$, we obtain near
the AS edge $\Omega \geq W_{s}$ the power law 
\begin{equation}
I(\Omega )\propto \left( \frac{\Omega -W_{s}}{W_{s}}\right) ^{\beta
}\,,\;\beta =\frac{v_{F}}{4u}  \label{beta}
\end{equation}
which is much more pronounced than the exponential law (see (\ref{gap}) below) for gapful cases.

Our derivation suggests literally the long range order at large $(x,t)$
distances and neglects fluctuations of the phase except perturbations
enforced by the instanton. 
But the mean fluctuations of the phase diverge and the order parameter
decays in a power law. These long range fluctuations are not related to the
instanton and can be taken into account {\it a posteriori. }
(Actually we have been using only the
discontinuity condition leaving the phase free at infinity.)It is easy to do
by noticing that the eigenfunctions in the prefactor of (\ref{I}) transform
as $\Psi _{0}\Rightarrow \Psi _{0}\exp \{i/2\,\varphi (x,t)\}$ which
averaging contributes the action term 
\[
\delta S_{\varphi }=1/8<[\varphi (0,0)-\varphi (0,T)]^{2}>\approx
\left(u/4v_{F}\right) \ln \left( \,uT/\xi _{0}\right) \,.
\]
Thus the effect of phase fluctuations, as well as the major role of the form
Factor, is simply to correct the value of the index in (\ref{beta}) as
$\beta\Rightarrow \beta ^{\ast }=v_{F}/4u+u/4v_{F}$. Within our adiabatic
approximation $u/v_{F}\ll 1$ the correction is small but it builds a bridge
to quantum nonadiabatic models where exactly the combination of
$\beta^{\ast }$ appears as the index of the single particle Green function with
$\gamma _{\rho }=u/v_{F}$ being identified as the charge channel exponent.
The link is completed by noticing that the AS is a realization of the spinon
and that the phase discontinuity in (\ref{s-snd}) is equivalent, together
with fluctuations, to applying the operator 
\[
\exp \left\{ \frac{i}{2}\,\varphi (x,t)+i\frac{\pi }{2}
\frac{\delta }{\delta\varphi }{\rm sgn}x\Theta (t)\Theta (T-t)\right\}
\]
which is our limit for the bosonization.

\section{Acoustic polaron and the free edge. \label{polaron}}

\subsection{ 1D semiconductors with acoustic and optical polarons.}

Behavior near the free edge $\Omega \approx \Delta _{0}$ is dominated by
small fluctuations $\eta $ of the gap amplitude $|\Delta |=\Delta _{0}+\eta $
and of the Fermi level $\delta E_{F}=\varphi ^{\prime }v_{F}/2$ via the
phase gradient $\varphi ^{\prime }=\partial _{x}\varphi $. We shall consider
it in a frame of a generic problem of the combined (gapful and acoustic)
polaron. The more simple, in compare to the CDW, single particle formulation
bares similar qualitative features but allows for a more detailed analysis.
Consider electron (hole) states in a 1D dielectric near the edge of a
conducting (valence) band. We shall take into account the gapful mode $\eta $
with the coupling $g_{0}$ and the sound mode (for which we shall keep the
''phase'' notation $\varphi$) with the velocity $u$ and the coupling
$g_{s}$. In generic semiconductors the sound mode is always present as the
usual acoustic phonon while the gapful mode can be present as an additional
degree of freedom. In all CDWs the gapful mode is always present as the
amplitude fluctuation $|\Delta| =\Delta _{0}+\eta$ while the sound mode
appears in the ICDWs as the phase $\Delta =|\Delta |\exp [i\varphi ]$.

Within the adiabatic approximation for the electron's wave function $\Psi $
the action $S$ (at the imaginary time) has the form 

\begin{eqnarray}
S=\int dx\int_{0}^{T}dt\left[ \left( \frac{1}{2m}\left| \partial _{x}\Psi
\right| ^{2}-\Omega \left| \Psi \right| ^{2}\right) +\left( g_{s}\partial
_{x}\varphi +g_{0}\eta \right) \Psi ^{\dagger }\Psi \right] +
\nonumber\\
\int dx\int_{-\infty }^{\infty }dt\left[ \frac{K_{s}}{2}\left( \left(
\partial _{t}\varphi /u\right) ^{2}+\left( \partial _{x}\varphi \right)
^{2}\right) +\frac{K_{0}}{2}\left( \left( \partial _{t}\eta /\omega
_{0}\right) ^{2}+\eta ^{2}\right) \right] \,. 
\label{S-pol}
\end{eqnarray}
Thus for the ICDW case we have $m=\Delta _{0}/v_{F}^{2}$, $g_{0}=1$,
$g_{s}=v_{F}/2$, $K_{s}=v_{F}/2\pi $, $K_{0}=4/\pi v_{F}$,
$2^{3/2}u/v_{F}=\omega _{0}/\Delta _{0}$ and $\Omega $ is counted with
respect to the edge $\Delta _{0}$ rather than to the middle of the gap is in
the previous chapter.

It is well known \cite{rashba}{\em \ }that the stationary state, the time
independent extremum of (\ref{S-pol}), corresponds to the selftrapped
complex, the polaron. Here it is composed equally by both $\eta $ and
$\varphi ^{\prime }$ which contribute additively to the static
coupling (while the dynamics will be completely different)

\[
\lambda =\lambda _{s}+\lambda _{0}=\frac{g_{s}^{2}}{K_{s}}+
\frac{g_{0}^{2}}{K_{0}}
\]
The polaronic length scale $l$ for $\eta \sim \varphi ^{\prime }\sim |\Psi
|^{2}\equiv \rho _{p}(x)$ is $l=2/m\lambda $ and the total energy is
$W_{p}=-m\lambda ^{2}/24$. The conditions $|W_{p}|\gg \omega _{0}$ and
$\lambda \gg u$ define the adiabatic, Born - Oppenheimer, approximation. For
the CDW case $\lambda _{s}=v_{F}\pi /2$ and $\lambda _{0}=v_{F}\pi /4$ hence 
$\lambda \sim v_{F}$ and we would arrive at $|W_{p}|\sim \Delta _{0}$ and
$l\sim \xi _{0}=v_{F}/\Delta _{0}$ which are the microscopic scales where the
single electronic model may be used only qualitatively. The full scale
approach for nearly stationary states has been considered above in
Ch.\ref{soliton}, but the upper PG region near the free edge $\Delta _{0}$ will be
described by the model (\ref{S-pol}) even quantitatively and most efficiently.

We can integrate out the fields $\varphi $ and $\eta $ at all $(x,t)$ to
obtain the action in terms of $\psi $ alone which is defined now only at the
interval $(0,T)$ for $t$: 
\begin{eqnarray}
S\{\Psi ;T\}=\int dxdt\left( \frac{1}{2m}\left| \partial _{x}\Psi \right|
^{2}-\Omega \left| \Psi \right| ^{2}\right) -
\frac{1}{2}\int \int
dt_{1,2}\int \int dx_{1,2} 
\nonumber\\
\left\{ U_{0}(x_{1}-x_{2},t_{1}-t_{2})\rho (x_{1},t_{1})\rho
(x_{2},t_{2})+U_{s}(x_{1}-x_{2},t_{1}-t_{2})\partial _{x}\rho
(x_{1},t_{1})\partial _{x}\rho (x_{2},t_{2})\right\}
\label{S-sT} 
\end{eqnarray}
Here the self-attraction retarded potentials are 
\begin{equation}
U_{s}=\frac{\lambda _{s}u}{2\pi }\ln \sqrt{x^{2}+t^{2}u^{2}}\,,\;U_{0}=
\frac{1}{2}\lambda _{0}\omega _{0}\exp [-\omega _{0}|t|]\delta (x)  \label{Us-U0}
\end{equation}

An equivalent form, suitable at large $T$, is obtained via integrating by
parts 
\begin{eqnarray}
S\{\Psi ;T)=\int dx\int_{0}^{T}dt\left[ \frac{1}{2m}\left| \partial _{x}\Psi
\right| ^{2}-\Omega \rho -\frac{\lambda }{2}\rho ^{2}\right] +  
\nonumber\\
\frac{1}{2}\int \int dt_{1,2}\int \int dx_{1,2}\partial _{t}\rho
(x_{1},t_{1})\partial _{t}\rho (x_{2},t_{1})U(x_{1}-x_{2},t_{1}-t_{2})
\label{S-lT} 
\end{eqnarray}
where $U(x,t)=u^{-2}U_{s}+\omega _{0}^{-2}U_{0}$.

The absorption near the absolute edge $\Omega \approx W_{p}$ is determined
by the long time processes when the lattice configuration is almost
statically self-consistent with electrons. The first term in (\ref{S-lT}) is
nothing but the action $S_{st}$ of the static polaron which extremum at
given $T$ is 
\[
S_{st}\approx -T\delta \Omega \,,\;\delta \Omega =\Omega -W_{p}\,.
\]
The second term in (\ref{S-lT}) $S_{tr}$ collects contributions only from
short transient processes near the impact moments $t=0,T$ which are seen by
the long length part as $\partial _{t}\rho (x,t)\approx \rho _{p}(x)[\delta
(t)-\delta (t-T)]$ where $\rho _{p}$ is the density for the static polaron
solution. We obtain 
\begin{eqnarray*}
S_{tr} &\approx &\int \int dx_{1,2}\rho _{p}(x_{1})\rho
_{p}(x_{2})U(x_{1}-x_{2},T) \\
&=&\frac{\lambda _{s}}{2\pi u}\ln \frac{uT}{l}+C_{0}
\frac{\lambda _{0}/l}{\omega _{0}}\exp [-\omega _{0}T]+cnst
\end{eqnarray*}
with $C_{0}\sim 1$. We see the dominant contribution of the sound mode which
grows logarithmically in $T$ while the part of the gapful mode decays
exponentially. If the sound mode is present at all, then the extremum over
$T$ is
\begin{equation}
T\approx \frac{\lambda _{s}}{2\pi u}\frac{1}{\delta \Omega }\,,\;
S\approx \frac{\lambda _{s}}{2\pi u}\ln \frac{C_{s}|W_{p}|}{\delta \Omega }
\,,\;C_{s}\approx 0.9  
\label{Cp}
\end{equation}
We find that near the absolute edge $\Omega \approx W_{p}$ the absorption is
dominated by the power law with an index $\alpha $ which must be big within
our adiabatic assumption $\alpha \gg 1$:
\begin{equation}
I\sim \left( \frac{\delta \Omega }{|W_{p}|}\right) ^{\alpha }\,,\;\alpha
=\frac{\lambda _{s}}{2\pi u}  \label{power}
\end{equation}
For parameters of the ICDW we obtain $\alpha =v_{F}/4u$ in full accordance
with the exact treatment (\ref{beta}).

Only in absence of sound modes $\lambda _{s}=0$ the gapful contribution can
determine the absolute edge. Then the minimization over $T$ of
$S=S_{core}+\delta S_{gap}$ would lead qualitatively to the result of
\cite{matv-01}:
\begin{equation}
T\sim \omega _{0}^{-1}\ln \left| \frac{W_{p}}{(W_{p}-\Omega )}\right|
\;,\;I\sim \exp \left( -cnst\frac{|W_{p}|}{\omega _{0}}+
\frac{\Omega -W_{p}}{\omega _{0}}
\ln \left| \frac{W_{p}}{(\Omega - W_{p} )}\right| \right)
\label{gap}
\end{equation}
for $\Omega \approx W_{p}$.

\subsection{Free edge vicinity.}

Consider the opposite regime near the free edge $\Omega \approx 0$
($\Omega\Rightarrow \Omega -\Delta _{0}$ for the ICDW). Here, entering the PG at
$\Omega <0$, the absorption is determined by fast processes of quantum
fluctuations: their characteristic time $T=T(\Omega )$ is short in compare
to the relevant phonon frequency: $T\ll \omega _{0},u/L$, where $L=L(\Omega )$ is the characteristic localization length for the fluctuational electronic
level at $E_{0}=\Omega $.  Since $T$ is small, we can neglect all variations
in time within $(0,T)$. Then we estimate the action (\ref{S-sT}), term by
term, as 
\begin{equation}
S\approx \frac{C_{1}T}{mL^{2}}-\Omega T-C_{2}\lambda _{s}u
\left( \frac{T}{L}\right) ^{2}-C_{3}\lambda _{0}\omega _{0}\frac{T^{2}}{L}  \label{cmb-sh}
\end{equation}
with $C_{i}\sim 1$. Its extremum over both $L$ and $T$ yields 
\[
S\sim \frac{|\Omega |^{3/2}/m^{1/2}}{\max
\left\{ |m\Omega |^{1/2}u\lambda_{s};\omega _{0}\lambda _{0}\right\} }
\]
which provides a reasonable interpolation  for the absorption in the closest
and the more distant vicinities of the free electronic edge. For the purely
acoustic case $\lambda _{0}=0$, a variational estimation for the numerical
coefficient as $C_{1}\approx 1/6,C_{2}\approx 0.06$ gives

\begin{equation}
I\sim \exp [-cnst|\Omega |/mu\lambda _{s}]\,,\;cnst\approx 2.8  \label{free}
\end{equation}
The validity condition $uT/L\,\sim \sqrt{-\Omega /W_{p}}\ll 1$ is satisfied
by definition of the edge region. This condition is compatible with the low
boundary for the frequency: $S\gg 1$ hence $-\Omega /W_{p}\gg u/\lambda _{s}$, which is small as our basic adiabatic parameter.

For gapful phonons alone, $\lambda _{s}=0$, we arrive at the known result 
(see \cite{matv-01} and references therein)
$I\sim \exp [-cnst|\Omega |^{3/2}/\omega _{0}]$, $\Omega <0$. But it was not
quite predictable that among the laws $S\sim |\Omega |^{3/2}$ and $S\sim
|\Omega |$ this is the smallest contribution to $S$ which wins: $\sim
|\Omega |^{3/2}$ at lowest $|\Omega |$ and $\sim |\Omega |$ for larger
$|\Omega |$. For the ICDW particularly, we have $\lambda _{0}/\lambda _{s}\sim 1$ and
$u/\omega _{0}\sim \xi _{0}$, then there is no space for the intermediate
asymptotics $\ln I\sim \Omega $ at $|\Omega |\ll \Delta _{0}$: beyond the
region with $S\sim |\Omega |^{3/2}$ dominated be the amplitude fluctuations,
the phase-only description is not valid and the particular nature of
amplitude solitons must be taken into account. This regime has been
considered above in Ch.\ref{soliton}.

Actually the difference between the laws $\log I\sim -|\Omega |/u$ and
$\log I\sim -|\Omega |^{3/2}/\omega _{0}$ can be easily interpreted. Indeed, for
gapful phonons we expect the frequency scale to be $\omega _{0}\Rightarrow
\omega _{k}=uk\sim u/L\sim u|\Omega |^{1/2}$ where $k\sim 1/L$ is a
characteristic wave number and $L$ is the localization length of the fluctuation
providing the bound state at $-\Omega $.
Then $|\Omega |^{3/2}/\omega_{0}\Rightarrow
|\Omega |^{3/2}/\omega _{k}\sim |\Omega |/u$.

While the law (\ref{free}) looks to be the simplest one, actually it is
quite uncommon and its derivation is problematic in all systems, cf. \cite
{iosel}. In our case we notice that only at $\lambda _{0}\neq 0$ the action
(\ref{cmb-sh}) has a usual saddle point: minimum over $L$ and maximum over $T$. But for the purely acoustic case $\lambda _{0}=0$ the minimum over $L$
appears only along the extremal line over $T$. Contrarily, at a given $T$
the action collapses to either $L\rightarrow 0$ or to $L\rightarrow \infty $
depending on a value of $T$ with respect to the threshold
$T^{\ast }\sim(mu\lambda _{s})^{-1}$ which is just the inverse width in (\ref{free}).
The paradox can be resolved inspecting the generic real time formulation
(\ref{I}). But a necessary insight will be obtained more easily by another treatment
presented in the next section.

\subsection{Quantum fluctuations as an instantaneous disorder with long rang
space correlations.\label{noise}}

It has been noticed already that in a one dimensional system the optical
absorption near the band edge can be viewed as if for a quenched disorder which
is emulated by instantaneous quantum fluctuations. This asymptotically exact
reduction to the time independent model can be done as follows. After
neglecting the retardation at $T\ll \omega _{0},u/L$, the self-interaction
term in (\ref{S-sT}) can be decoupled back by the Hubbard-Stratonovich
transformation via the {\em time independent }field $\zeta $ with the
correlator $D(x)=U_{0}(x,0)+\partial _{x}^{2}U_{s}(x,0)$:

\begin{equation}
S\{\Psi ,\zeta ;T)=T\int dx\left( \frac{1}{2m}\left| \partial _{x}\Psi
\right| ^{2}+\zeta (x)\rho (x)\right) +\frac{1}{2}\int \int dx_{1,2}\zeta
(x_{1})D^{-1}(x_{1}-x_{2})\zeta (x_{2})  \label{S-psi-xi}
\end{equation}
After integration over $\Psi $ and rotation to the real time it becomes
finally the DOS expression 
\[
\int D[\zeta (x)]\delta (E[\zeta (x)]-\Omega )\exp \left[ -\frac{1}{2}\int
\int dx_{1}dx_{2}\zeta (x_{1})D^{-1}(x_{1}-x_{2})\zeta (x_{2})\right] 
\]
Here $E[\zeta (x)]$ is the eigenfunction in the random field $\zeta $: 
\[
\frac{-\partial _{x}^{2}}{2m}\Psi +\zeta \Psi =E\Psi \,.
\]
For the dispersionless phonon alone, e.g. the amplitude mode in the CDW,
$D(x)=U_{0}(x,0)\sim \delta (x)$ and the known exact results for the
uncorrelated disorder \cite{disorder} provide us with the PG
asymptotics 
\begin{equation}
I(\Omega )\propto \exp \left[ -\frac{8}{3^{3/2}}
\frac{\left| W_{p}\right| }{\omega _{0}}\left|
\frac{\Omega }{W_{p}}\right| ^{3/2}\right] \,.  \label{Ig}
\end{equation}
For the CDW parameters it becomes

\begin{equation}
I(\Omega )\propto \exp \left[ -\frac{16}{3\pi }\left( 2\frac{\Delta
_{0}-\Omega }{(\Delta _{0}\omega _{0}^{2})^{1/3}}\right) ^{3/2}\right] \,.
\label{Ig-cdw}
\end{equation}

Below we shall concentrate only on a more problematic case of the sound
mode. The correlator of the ''disordered potential'' $\zeta $, $D(x)$, is
just the mean square of quantum fluctuations of the phonon potential
$\zeta=v_{F}/2\varphi ^{\prime }(x,t)$ at coinciding times: in the Fourier
representation 
\[
D_{k}=\int \frac{d\omega }{2\pi }\frac{\lambda _{s}k^{2}}
{(\omega/u)^{2}+k^{2}}=\frac{1}{2}\lambda _{s}u|k| \,.
\]
The probability distribution for the Fourier components $\zeta _{k}$ is 
\begin{equation}
P[\zeta _{k}]\sim \exp \left( -\int \frac{dk}{2\newline
\pi }\frac{|\zeta _{k}|^{2}}{\lambda _{s}u|k|}\right)   \label{zeta-k}
\end{equation}
which tells us that the component with $k=0$ is excluded, $P(\zeta _{0})=0$.
The constraint 
\begin{equation}
\zeta _{0}=\int_{-\infty }^{\infty }\zeta (x)dx=0\,.  \label{zero}
\end{equation}
agrees with properties of the potential $\sim \varphi ^{\prime }$ in the
time dependent picture of the previous section which obeys the condition
(\ref{zero}) at any finite $t$, see the  Fig.3. Contrary to usual
expectations for the method of optimal fluctuations, here the potential well
creating the level $E$ must be accompanied by compensating barriers. The
condition (\ref{zero}) is linked to the paradox noticed in the pervious
section: an absence of a finite minimum over the length scale at a given $T$. Indeed, now we cannot rely upon an existence of a bound state at arbitrary
shallow potential: $E_{0}\sim -m\left( \int dx\zeta (x)\right) ^{2}$ which
is zero at the condition (\ref{zero}).

While the divergency at small $k$ (large $x$) is physical, the one at large
$k$ in (\ref{zeta-k}) must be regularized to work in the real space. We shall
proceed by introducing the auxiliary field $\mu (x)$ such that
$\zeta =d\mu/dx=\mu ^{\prime }$. Finally we arrive at the model of the ''nonlocal
acoustic disorder'' 
\begin{equation}
I(\Omega )\sim \int D[\mu (x)]\delta (E[\partial _{x}\mu (x)]-\Omega )\exp 
\left[ -\frac{\lambda _{s}}{2u}\int \int dx_{1}dx_{2}\frac{\left( \mu
(x_{1})-\mu (x_{2})\right) ^{2}}{|x_{1}-x_{2}|^{2}}\right]   \label{mu-mu}
\end{equation}
Here the integral in the exponent is already regular at small $x$. The
divergence at large $x$ works to maintain the constraint (\ref{zero}),
otherwise $\mu (+\infty )-\mu (-\infty )=\int_{-\infty }^{\infty }\zeta
(x)dx\neq 0$ and the integral in (\ref{mu-mu}) would diverge logarithmically
leading to the zero probability.

Unfortunately we are not aware of exact studies for disordered systems with
such long range correlations. Usual scaling estimations (\cite{lifshits}) for
characteristic $\mu $ and $l$ give us 
$|\Omega| \sim 1/(ml^{2})\sim |\mu |/l$
then $|\mu| \sim |\Omega /m|^{1/2}$ hence $\ln I\sim -\mu ^{2}\lambda
_{s}/u\sim -|\Omega |\lambda _{s}/u$ in accordance with direct estimations
and the result (\ref{free}) for the generic time dependent model.

\section{Discussion and Conclusions}

We summarize the obtained results as follows.

The PG starts below the free edge by (stretched) exponential dependencies 
\begin{equation}
I\sim \exp \left[ -cnst\left( -|\Omega |\right) ^{\gamma }\right] \,
\label{edge}
\end{equation}
with different powers $\gamma =3/2$ for gapful phonons and $\gamma =1$ for
sounds. If both modes are present, then the smallest one, with $\gamma =3/2$, dominates at small $\Omega $. This regime corresponds to free electronic
states smeared by instantaneous uncorrelated quantum fluctuations of the
lattice.

Deeply within the PG, approaching the absolute threshold $W_{s}$ or $W_{p}$,
the exponential law changes for the power law
$I(\Omega )\propto \left(\Omega -W_{s}\right) ^{\beta }\,$with the big index $\beta $.
This contribution dominates over the smooth one from gapful modes
$I\sim \exp(cnst\,\delta \Omega \ln \delta \Omega )$. The power regime corresponds to
creation of nearly amplitude solitons surmounted be compensating phase
tails. Its description provides a quasiclassical interpretation for
processes in fully quantum systems of correlated electrons in the spin-gap
regime, with the AS being a version of the spinon.

These results are different from anything used earlier both in theoretical
discussions and in interpretation of experimental data \cite{ohio}. They can
vaguely explain the unusually wide pseudogaps observed in experiments even
at low temperatures for well formed ICDWs.

Our results have been derived for single electronic transitions: PES and tunneling. They can also be applied to intergap (particle-hole) optical transitions as long as semiconductors are concerned. For the ICDW results are applied to the free edge vicinity. But the edge at $2E_s$ will disappear in favor of optically active gapless phase mode.

It should be stresses in this respect that there {\em cannot be a common PG }
for processes characterized by different time scales. We should distinguish 
\cite{braz-80} between short living states observed in optical, PES (and may
be tunneling) experiments and long living states (amplitude solitons, phase
solitons) contributing to the spin susceptibility, NMR relaxation, heat
capacitance, conductivity, etc. States forming the optical PG are created
instantaneously; particularly near the free edge they are tested over times
which are shorter than the inverse phonon frequencies
$\tau _{opt}\sim \hbar/E_{g}<\omega _{ph}^{-1}$ and many orders of magnitude beyond the life times
required for current carriers, and even much longer times for thermodynamic
contributions. It follows then that the analysis of different groups of
experimental data within the same picture \cite{ohio} must be reevaluated.
The lack of discriminating different time scales concerns also typical
discussions of PGs in $High-T_{c}$ superconductors.

We conclude that the subgap absorption in systems with gapless phonons is
dominated by formation of long space-time tails of relaxation. It concerns
both acoustic polarons in $1D$ semiconductors and solitons in CDWs. Near the
free edge the simple exponential, Urbach type, law appears competing with
stretched exponential ones typical for tails from optimal fluctuations. The
deeper part of the PG is dominated by the power law singularity near the
absolute edge.

\acknowledgments
S. M. acknowledges the hospitality of the Laboratoire de Physique Theorique
et des Modeles Statistiques, Orsay and the support of the CNRS and the ENS -
Landau foundation. This work was partly performed within the INTAS grant
\#2212.
\newpage
\begin{figure}[tbph]
\begin{center}
\includegraphics[width=3.0in]{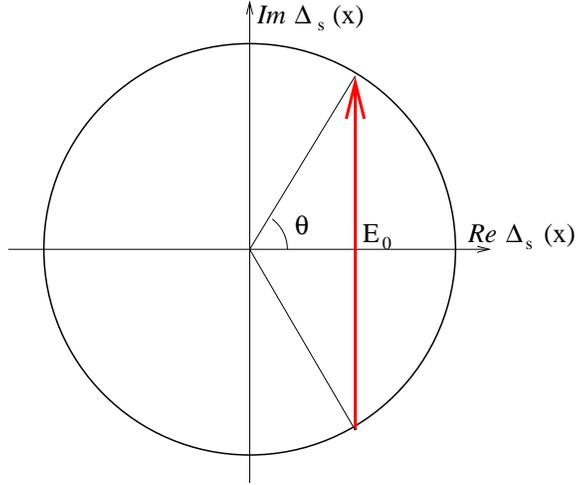}
\caption{Trajectory of the chordus soliton with phase tails in the complex
plane $\Delta $}
\end{center}
\label{chord}
\end{figure}

\begin{figure}[tbph]
\begin{center}
\includegraphics[width=3.0in]{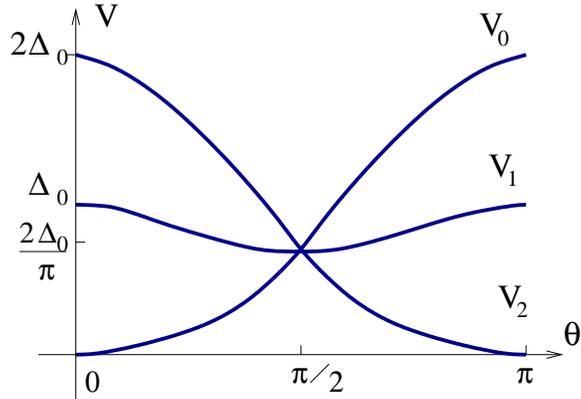}
\caption{Selftrapping terms $V_{\nu}$ for chordus solitons
as functions of the chiral angle $2\theta$ for various fillings $\nu$.}
\end{center}
\label{terms}
\end{figure}

\begin{figure}[tbph]
\begin{center}
\includegraphics[width=3.0in]{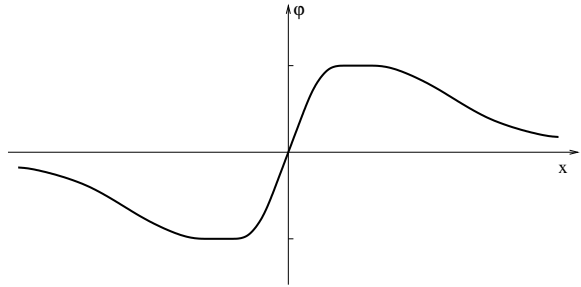}
\caption{The acoustic polaron field $\protect\varphi (x,t)$ as a function of 
$x$ at some moment $t$.}
\end{center}
\label{phase}
\end{figure}

\newpage
\section{Appendix}

Electronic energies in the complex field $\Delta $ are determined by the
Dirac Hamiltonian 
\[
\left| 
\begin{array}{cc}
-iv_{F}\partial _{x} & \Delta \\ 
\Delta ^{\ast } & iv_{F}\partial _{x}
\end{array}
\right| \, , \,
\Delta =|\Delta |e^{i\varphi}
\]
In the ground state $|\Delta (x,t)|=\Delta _{0}$, $\varphi =cnst$ and the
electronic spectrum is $E^{2}=v_{F}^{2}k^{2}+\Delta _{0}^{2}$ where $v_{F}$
is the Fermi velocity. But these free states are not proper excitations. The
evolution of added electrons or holes with the initial $E_{0}\ge\Delta _{0}$
to the AS with $W_{s}=2/\pi \Delta _{0}<\Delta _{0}$ can be described by the
exact solution for intermediate configurations characterized by the singly
occupied arbitrary positioned intragap state $E_{0}=\Delta _{0}\cos \theta $
with $0\leq \theta \leq \pi $, hence $-\Delta _{0}<E_{0}<\Delta _{0}$. It
was found \cite{braz-80} to be the Chordus Soliton ($ChS$) with $2\theta $
being the total chiral angle: $\Delta (\pm \infty )=\exp (\pm i\theta )$,
see the Fig.1.

Namely, 
\begin{equation}
\Delta _{ChS}(x,\theta )=\Delta _{0}(\cos \theta +
i\sin \theta \tanh (k_{0}x))\exp i\varphi _{0},\quad
k_{0}=\Delta _{0}\sin \theta .
\end{equation}
with an arbitrary $\varphi _{0}=cnst$. The potentials $V_{\nu }$ are known 
\cite{braz-80} as  (see Fig.2
\[
V_{\nu }(\theta )=\Delta _{0}[(\nu -\frac{2}{\pi }\theta )\cos \theta
+\frac{2}{\pi }\sin \theta ]\,
\]
where $\nu $ is the filling number of the intragap state that is $\nu =0,1$
for $j=0,1$ while $\nu =2$ is equivalent to $j=0$ for the ground state
extended by the two particles $N=2M+2$. The term $V_{0}(\theta )$ increases
monotonically from $V_{0}(0)=0$ for the $2M$ ground state to $V_{0}(\pi
)=2\Delta _{0}$ for the $2M+2$ ground state (GS) with two free holes.
Apparently there is an opposite dependence for
$V_{2}(\theta )=V_{0}(\pi-\theta )$.
Thus the total phases slip $2\theta =0\Rightarrow 2\theta =2\pi $
realizes the spectral flow across the gap, accompanied also by the flow of
particles for $\nu =2$ which makes it favorable. The term $V_{1}(\theta
)=V_{1}(\pi -\theta )$ is symmetric describing both the particle upon the
$2M$ GS and the hole upon the $2M+2$ GS. Apparently
$V_{1}(0)=V_{1}(\pi)=\Delta _{0}$
(the degenerate Ground States are: the $2M$ one with the
additional free electron for $\theta =0$ and the $2M+2$ one with the
additional free hole for $\theta =\pi $) while the minimum is
$V_{1}(\pi/2)=W_{s}<\Delta _{0}$ where $W_{s}=2\Delta _{0}/\pi $ is the AS energy.
Thus the stationary state of the system with an odd number of particles, the
minimum of $V_{1}$, is the amplitude soliton ($AS$)
$\Delta \Rightarrow-\Delta $ with the midgap state $E_{0}=0$ occupied by the singe electron.

Notice that being the uncharged spin carrier with the topological
charge equal unity, the AS is a quasiclassical realization of the spinon in
systems with nonretarded attraction of electrons (that is with high, rather
than low, phonon frequencies).
\end{document}